\documentclass[10pt,twocolumn]{article}

\usepackage[T1]{fontenc}
\usepackage[margin=1in]{geometry}
\usepackage{graphicx}
\usepackage{amsmath}
\usepackage{amssymb}
\usepackage{booktabs}
\usepackage{tabularx}
\usepackage[table]{xcolor}
\usepackage{array}
\usepackage{microtype}
\usepackage{hyperref}
\usepackage{enumitem}
\usepackage{titlesec}
\usepackage{abstract}

\titlespacing*{\section}{0pt}{8pt}{4pt}
\titlespacing*{\subsection}{0pt}{6pt}{3pt}

\definecolor{capGood}{HTML}{DFF2D8}
\definecolor{capMid}{HTML}{FFF0CC}
\definecolor{capBad}{HTML}{FADBD8}
\definecolor{capHdr}{HTML}{F2F2F2}

\title{\textbf{Data Facts: A Metadata Schema for Structured Data Exchange in the NANDini Multi-Agent Ecosystem}}

\author{Jin Gao, Maria Gorskikh, Pradyumna Chari, Brittany Box,\\
Mukul Kemla, Pratik Behera, Abhishek Mehta, Ramesh Raskar}

\date{}

\begin{document}

\maketitle

\begin{abstract}
NANDini (Networked Agents Natural Distillation of Interconnected Nodal Intelligence) envisions an automated ecosystem where intelligent agents independently create, process, and exchange data to drive decisions at scale. Realizing this vision requires infrastructure beyond agent discovery and communication: agents must be able to advertise, evaluate, and verify the datasets they hold. Current protocols (NANDA for federated registry, A2A and MCP for inter-agent messaging) address identity and communication but provide no mechanism for structured data exchange. Existing Enterprise data-sharing frameworks (IDS-RAM, Gaia-X, Ocean Protocol) assume human-in-the-loop governance incompatible with autonomous, real-time agent interactions. We introduce \textbf{Data Facts}, a core NANDini concept: a lightweight JSON metadata schema that bridges agent discovery and data access via a single pointer (\texttt{data\_facts\_url}) added to an existing Agent Facts registry record. The linked document encodes dataset identity, access tier (public, semi-private, or private), endpoint, a time-to-live for freshness validation, and a SHA-256 integrity checksum. For private and semi-private data, we implement a three-layer security pipeline: JWT authentication, capability-scoped gateway authorization, and an A2A credential delegation protocol. Across 840 decision-making evaluations, data-informed agents achieve 100\% accuracy versus 35.2\% without data access ($p < 0.001$); TTL enforcement reduces stale-data errors from 37.6\% to 8.8\%; checksum verification achieves 100\% corruption detection at all injection rates; and the security pipeline blocks all 46 forgery attempts with zero data leakage.
\end{abstract}

\section{Introduction}

In a world of autonomous agents numbering in the hundreds of billions, the data-to-decision lifecycle must evolve into a continuous, distributed process. NANDini (Networked Agents Natural Distillation of Interconnected Nodal Intelligence)~\cite{nandini2025} envisions a planetary-scale ecosystem in which agents generate data, refine information, extract insights, and take action in response to human needs. Realizing this vision requires infrastructure that goes beyond agent identity and communication to include a structured data layer through which agents can exchange, interpret, and distill information across large-scale networks.

Current agent infrastructure addresses discovery and messaging. The NANDA Index~\cite{raskar2025nanda} provides a federated registry in which agents publish Agent Facts documents encoding identity, endpoints, and capabilities. Protocols such as Google A2A~\cite{google2025a2a} and Anthropic MCP~\cite{anthropic2024mcp} standardize inter-agent messaging and tool invocation across vendors. Together, these mechanisms enable agents to discover each other and exchange tasks. However, they provide no mechanism for advertising, evaluating, or verifying the datasets held by participating agents. An agent seeking data from a peer must rely on hard-coded endpoint knowledge or ad-hoc probing; no machine-readable layer exists to describe what a dataset contains, when it was last refreshed, whether it has been tampered with, or what credentials are required.

Prior work addresses data exchange at the organizational level. IDS-RAM~\cite{ids2019}, Gaia-X~\cite{gaiax2021}, and the Dataspace Protocol~\cite{dsp2023} define governance frameworks for enterprise-to-enterprise sharing; Ocean Protocol~\cite{ocean2022} introduces a blockchain-based marketplace for data NFTs. These systems assume human-in-the-loop governance, heavyweight connector infrastructure, and negotiation cycles that do not compose with the lightweight, autonomous agent interactions that NANDini targets. W3C DCAT~\cite{w3cdcat}, PROV~\cite{w3cprov}, and DID~\cite{w3cdid} provide vocabulary for cataloging, provenance, and identity but no runtime mechanism for agents to act on that vocabulary at query time. Table~\ref{tab:positioning} positions Data Facts against this prior work.

We propose \textbf{Data Facts}, a core concept within the NANDini framework: a JSON metadata schema that extends Agent Facts with a single new field, \texttt{data\_facts\_url}, pointing to an externally hosted document describing a dataset. The design maintains a strict separation of concerns. Agent Facts retain their existing lightweight registry role as a registry for agent identity, endpoints, and capabilities, while dataset metadata evolves independently and is resolved only when data is requested. For private data, access is protected through a three-layer security pipeline consisting of JWT-based authentication, capability-scoped authorization enforced through a mediating gateway, and an A2A negotiation protocol for scoped, time-bound credential delegation. By standardizing how agents describe and verify datasets, Data Facts operationalizes NANDini's commitment to data accuracy within autonomous multi-agent ecosystems.

\smallskip
\noindent\textbf{Evaluation scope.}
We report five experiments: discovery overhead, freshness enforcement, integrity verification, security pipeline evaluation, and data-informed decision quality. Decision-making experiments use counter-intuitive business queries designed to expose agents that reason from training priors rather than querying live data. The security evaluation examines 13 attack categories. We do not claim the schema is optimal for all agent architectures, nor that the security pipeline addresses every possible threat model.

\smallskip
\noindent\textbf{Contributions:}
\begin{enumerate}[leftmargin=*, itemsep=2pt, topsep=2pt]
    \item \textit{NANDini data layer:} Data Facts, a core NANDini concept providing a lightweight JSON metadata schema for agent-held datasets, integrating with Agent Facts via a single URL pointer with three-tier access semantics (\texttt{"public"}, \texttt{"semi\_private"}, \texttt{"private"}).
    \item \textit{Security pipeline:} JWT authentication, capability-scoped gateway authorization, and an A2A credential delegation protocol for multi-agent private data access.
    \item \textit{Empirical validation:} 840 decision-making evaluations showing 100\% data-informed accuracy versus 35.2\% without data access; freshness and integrity experiments; and adversarial security evaluation blocking all 46 forgery attempts across 206 private-access trials.
\end{enumerate}

\section{The NANDini Initiative}
\label{sec:nandini}

NANDini (Networked Agents Natural Distillation of Interconnected Nodal Intelligence)~\cite{nandini2025} is an open research initiative addressing a fundamental question: in a future dominated by intelligent agents, what will happen to the data-to-decision life-cycle? The initiative responds to the emergence of agentic systems of intelligence, where agents serve not only as task executors but as the primary source of data generation, curation, and consumption.

NANDini builds upon MIT's NANDA architecture~\cite{raskar2025nanda}, extending early registry and discovery frameworks into a comprehensive ecosystem for autonomous agent data exchange. While NANDA provides the foundational layer for federated agent discovery via AgentFacts, NANDini extends this foundation to address three interconnected research domains: \textit{knowledge distillation} across agent networks, \textit{model context protocols} for structured access to tools and data, and \textit{large-scale agentic systems of intelligence} capable of operating autonomously at scale.

Central to the NANDini vision is the principle that reliable decision-making in autonomous systems depends on accurate and verifiable data. This principle motivates the initiative's core technical contributions, including \textbf{Data Facts}: a standardized mechanism through which agents can advertise, evaluate, and verify the datasets they hold. Data Facts bridges the gap between agent discovery (addressed by NANDA) and agent communication (addressed by A2A and MCP) by introducing a structured data exchange layer that enables agents to make informed, data-driven decisions autonomously.

The initiative is committed to being open, transparent, and capable of operating at planetary scale, targeting an emerging landscape in which unprecedented investment in AI infrastructure intersects with the \$1.3 trillion enterprise software market through autonomous agent ecosystems.

\section{Related Work}
\label{sec:related}

\begin{table*}[t]
\centering
\caption{Comparison of related systems across four properties
         relevant to autonomous agent data exchange.
         \checkmark~= fully supported;
         $\circ$~= partial or vocabulary only;
         ---~= not addressed.}
\label{tab:positioning}
\small
\setlength{\tabcolsep}{5pt}
\begin{tabular}{lcccc}
\toprule
\textbf{System} &
\textbf{\shortstack{Dataset\\discovery}} &
\textbf{\shortstack{Freshness\\signaling}} &
\textbf{\shortstack{Integrity\\verification}} &
\textbf{\shortstack{Tiered\\access control}} \\
\midrule
NANDA AgentFacts~\cite{raskar2025nanda}   & ---          & ---          & ---          & ---          \\
A2A Agent Cards~\cite{google2025a2a}      & ---          & ---          & ---          & ---          \\
MCP \texttt{mcp.json}~\cite{anthropic2024mcp} & ---      & ---          & ---          & ---          \\
IDS-RAM~\cite{ids2019}                    & $\circ$      & ---          & $\circ$      & \checkmark   \\
Gaia-X / DSP~\cite{gaiax2021,dsp2023}    & $\circ$      & ---          & ---          & \checkmark   \\
Ocean Protocol~\cite{ocean2022}           & $\circ$      & ---          & \checkmark   & \checkmark   \\
W3C DCAT~\cite{w3cdcat}                   & \checkmark   & $\circ$      & ---          & ---          \\
W3C PROV~\cite{w3cprov}                   & ---          & $\circ$      & ---          & ---          \\
\midrule
\textbf{NANDini Data Facts (ours)}         & \checkmark   & \checkmark   & \checkmark   & \checkmark   \\
\bottomrule
\end{tabular}
\end{table*}

\noindent\textbf{Agent infrastructure.}
The NANDA Index~\cite{raskar2025nanda}, which serves as NANDini's foundational discovery layer, provides a federated registry where agents register with AgentFacts documents encoding identity, endpoints, and capabilities, but no dataset state. A survey of registry designs~\cite{singh2025survey} confirms that NANDA AgentFacts, A2A Agent Cards, and MCP \texttt{mcp.json} descriptors all omit dataset-level metadata. Google A2A~\cite{google2025a2a} and Anthropic MCP~\cite{anthropic2024mcp} standardize inter-agent communication and tool invocation respectively; a survey of these protocols~\cite{hussain2025interop} notes that all address communication while leaving data description to application logic. Data Facts, as a NANDini core concept, fills this gap with a single \texttt{data\_facts\_url} pointer in AgentFacts, resolving dataset metadata externally without inflating the registry.

\noindent\textbf{Data exchange and cataloging.}
Enterprise frameworks define richer metadata but assume organizational governance. IDS-RAM~\cite{ids2019}, Gaia-X~\cite{gaiax2021}, and DSP~\cite{dsp2023} target sovereign inter-organizational sharing via certified connectors and negotiation cycles; Ocean Protocol~\cite{ocean2022} introduces blockchain-based data NFTs with Compute-to-Data. None compose with lightweight, real-time agent interactions. W3C DCAT~\cite{w3cdcat} provides RDF vocabulary for federated catalog discovery; W3C PROV~\cite{w3cprov} covers data lineage; W3C DID~\cite{w3cdid} underpins our JWT authentication pipeline. These standards supply rich catalog vocabulary but do not define machine-enforceable freshness or integrity fields actionable at agent query time.

\noindent\textbf{Data valuation and quality.}
Ghorbani and Zou~\cite{ghorbani2019datashapley} formalize Data Shapley for equitable attribution of training data value; extensions address federated settings~\cite{wei2020fedshapley} and market pricing~\cite{zhang2024datamarkets}. Complementary lines of work develop task-agnostic valuation via statistical divergence between buyer and seller distributions~\cite{amiri2023taskagnostic,lu2024privatemeasurements} and federated data acquisition without labeled validation sets~\cite{lu2024daved}. All of this literature assumes data is already accessible and prices or selects contributions post-hoc. Batini et al.~\cite{batini2009dq} survey data quality dimensions including timeliness and integrity as core axes. Data Facts externalizes two of these, freshness via \texttt{ttl\_seconds} and integrity via \texttt{checksum\_sha256}, as declarative fields agents verify before ingestion, instantiating data quality primitives at the discovery layer rather than in governance pipelines.


\section{Agentic Data Infrastructure}

\begin{figure}
    \centering
    \includegraphics[width=1\linewidth]{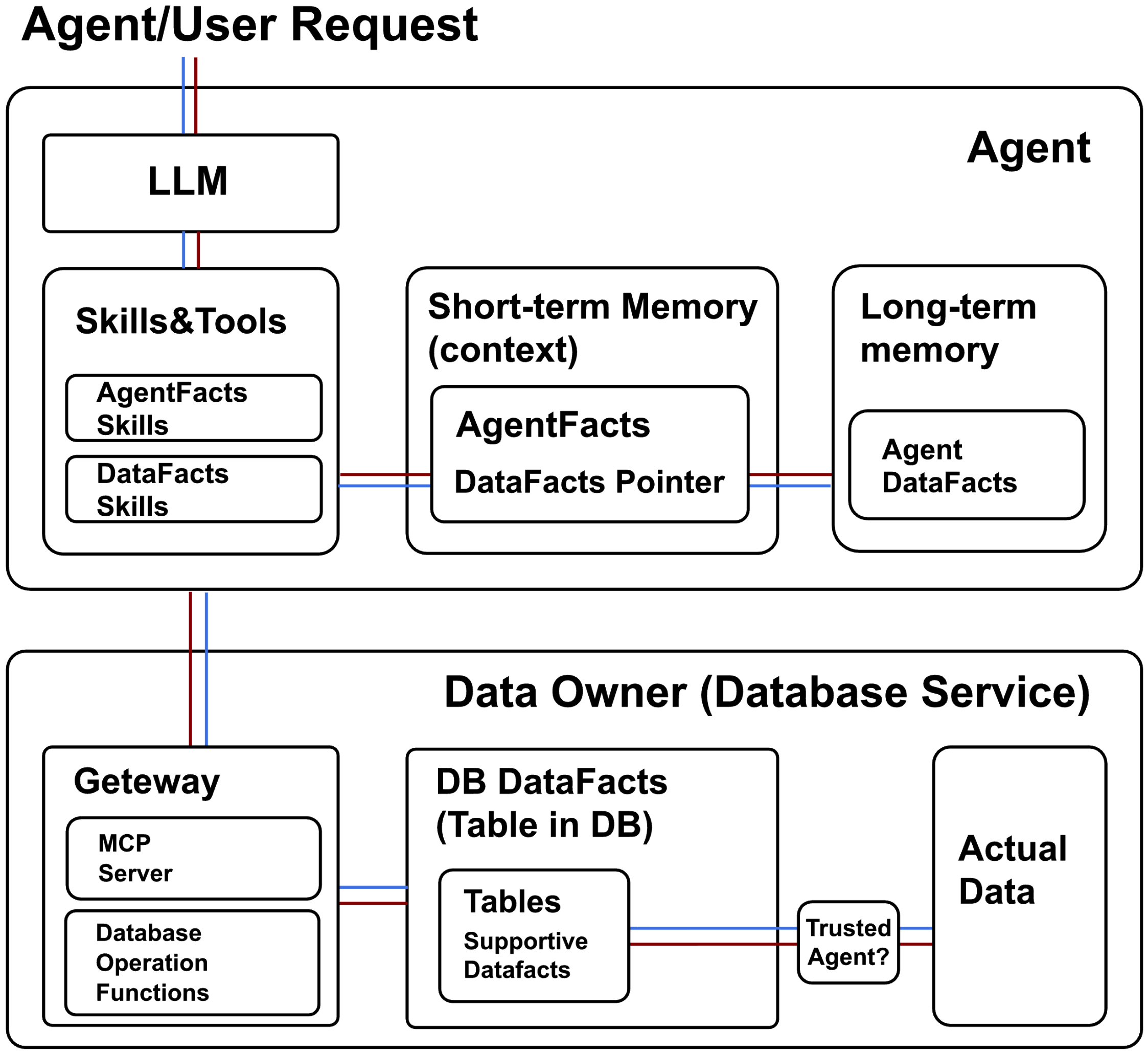}
    \caption{NANDini Data Facts architecture: agent layer (discovery, communication) and data-owner layer (gateway enforcement, capability delegation).}
    \label{fig:architecture-overview}
\end{figure}

Figure~\ref{fig:architecture-overview} shows the NANDini Data Facts architecture as two coordinated layers: an \textit{agent layer} and a \textit{data-owner layer}. User requests are interpreted by the LLM and executed
through skills/tools. Agents expose identity and endpoint metadata through \texttt{AgentFacts} (via the NANDA registry), while dataset access metadata is exposed through a \texttt{DataFacts} pointer (the NANDini data exchange layer).

On the data-owner side, each provider runs a gateway service in front of its database, together with control-plane tables for confidential access state, including \texttt{grants},
\texttt{delegations}, \texttt{agent\_keys}, and \texttt{audit\_log}. The gateway is the trust boundary: it verifies tokens, checks delegated scope and trust conditions, validates inputs, and
only then allows access to actual data.

Overall, the architecture realizes the NANDini vision of minimal, decentralized infrastructure for agent-driven data access by separating discovery, authorization, and execution, thereby enabling secure access to data and reliable exchange across multi-agent systems.
\subsection{Data Facts Schema Design}
\label{sec:design}

\noindent\textbf{Integration with AgentFacts.}
Data Facts extends the NANDA AgentFacts record with a single optional field, \texttt{data\_facts\_url}, which points to an externally hosted JSON document describing the agent's dataset.
The registry stores only a pointer, while the metadata document is resolved on demand.
This design preserves the registry's lightweight structure: the existing \texttt{/register} and \texttt{/list} endpoints require no structural changes, and agents that do not expose datasets incur no additional overhead.

A consumer agent's discovery workflow therefore proceeds as follows:
(1) query the registry (\texttt{GET /list}),
(2) extract \texttt{data\_facts\_url} from the returned AgentFacts record,
(3) fetch the Data Facts document over HTTP(S),
(4) validate freshness and integrity locally, and
(5) access the dataset endpoint.
Steps 3--5 are performed by the consumer without further interaction with the registry.

\noindent\textbf{Schema.}

The Data Facts document is a lightweight JSON object with a small \texttt{evidence} subobject.
Table~\ref{tab:schema} lists all fields.
Required fields cover the four properties identified in Section~\ref{sec:related}: dataset identity (\texttt{dataset\_id}, \texttt{dataset\_description}), access tier (\texttt{access\_type}), the resolvable dataset endpoint (\texttt{endpoint}), freshness (\texttt{ttl\_seconds}, \texttt{evidence.last\_updated}), and integrity (\texttt{evidence.checksum\_sha256}).
Optional fields (\texttt{evidence.source}, \texttt{update\_frequency}, \texttt{data\_owner}) carry provenance and human-readable cadence information.

\begin{table*}[t]
\centering
\caption{Data Facts schema fields.}
\label{tab:schema}
\small
\setlength{\tabcolsep}{4pt}
\begin{tabular}{lp{5.8cm}}
\toprule
\textbf{Field} & \textbf{Semantics and constraints} \\
\midrule
\multicolumn{2}{l}{\textit{Required}} \\[2pt]
\texttt{dataset\_id}               & Stable identifier for the dataset \\
\texttt{dataset\_description}      & Human-readable description \\
\texttt{access\_type}              & \texttt{"public"} $|$ \texttt{"semi\_private"} $|$ \texttt{"private"} \\
\texttt{endpoint}                  & HTTP(S) \textit{gateway} URL where data is served (never the producer's raw endpoint) \\
\texttt{evidence.last\_updated}    & ISO~8601 timestamp of last data update \\
\texttt{evidence.checksum\_sha256} & SHA-256 hex digest of the dataset (64 chars) \\
\texttt{ttl\_seconds}              & Cache validity window; must be $> 0$ \\
\midrule
\multicolumn{2}{l}{\textit{Optional}} \\[2pt]
\texttt{evidence.source}           & Provenance label (e.g.\ API name) \\
\texttt{update\_frequency}         & Human-readable update cadence \\
\texttt{data\_owner}               & Agent identifier of the data provider \\
\bottomrule
\end{tabular}
\end{table*}

\noindent\textbf{Freshness Enforcement.}

A consumer agent computes staleness as:
\[
  \Delta t = t_{\text{now}} - t_{\text{last\_updated}}
\]
and rejects the dataset if $\Delta t > \texttt{ttl\_seconds}$.
The TTL is set by the producing agent to reflect its actual update cadence; a finance feed agent updating every ten minutes would set \texttt{ttl\_seconds}~$= 600$.
This check is performed client-side from the Data Facts document, requiring neither a round-trip to the producer nor a query to the registry.


\noindent\textbf{Integrity Verification.}

The \texttt{evidence.checksum\_sha256} field contains the SHA-256 digest of the dataset at the time of the last update.
After fetching data from \texttt{endpoint}, the consumer recomputes the digest and compares it against the stored value.
A mismatch indicates either corruption in transit or a stale checksum resulting from an update cycle that has not yet refreshed the Data Facts document. In both cases the consumer can reject the data or re-fetch.

\noindent\textbf{Access Control.}

Semi-private capabilities are discoverable; private are not.
When \texttt{access\_type} is \texttt{"public"}, the consumer accesses \texttt{endpoint} directly.
When \texttt{access\_type} is \texttt{"semi\_private"} or \texttt{"private"}, access follows a three-layer pipeline: (i) the consumer presents credentials to obtain a JWT scoped to the requested dataset capability, (ii) the JWT is forwarded to a capability-scoped gateway that enforces per-dataset ACLs, and (iii) the gateway proxies requests; \texttt{endpoint} always points to the gateway.

\subsection{Gateway and Database-Side Infrastructure}
\label{sec:gateway-db}

The Data Layer Gateway serves as the enforcement boundary between agents and heterogeneous storage engines. Rather than exposing database-native protocols directly, each backend is fronted by a gateway service as an adapter layer. This separation keeps agent-to-agent negotiation and capability semantics stable even when execution targets differ across relational, document, object, vector, cache, search, graph, or time-series systems.

\noindent\textbf{Gateway behavior and trust enforcement.}
A request entering the gateway is processed through a single verification path before any data operation is executed. The gateway validates JWT signatures against registered public keys, extracts capability and operation claims, and checks expiry and revocation state. For delegated credentials, it also verifies delegation lineage, confirms that the underlying grant remains active, and enforces scope consistency so delegated operations cannot exceed delegated rights. In credential-pass mode, token binding is enforced by matching caller identity to the credential binding claim, preventing credential reuse by unintended agents.

After identity and delegation checks, the gateway applies replay protection through JTI tracking, optional per-agent rate limiting, and operation-level authorization against active capability grants. Request arguments are then validated against traversal and injection patterns and constrained by backend-specific scope guards such as bucket or collection allow-lists. Only after these controls have been satisfied does the gateway dispatch the request to the backend adapter and persist an audit record containing principal, capability, action, success state, and result cardinality.

\noindent\textbf{Execution paths.}
The architecture supports two operational paths with the same enforcement model. In A2A proxy mode, a requester sends a negotiated query to the data owner, and the owner executes locally before returning results over A2A. In credential-pass mode, the requester presents a delegated credential directly to the owner gateway at \texttt{/query}. Although the transport differs, authorization, validation, and auditing are enforced in the same gateway pipeline.

\noindent\textbf{Database-side control plane.}
Security state is materialized in control-plane tables, including \texttt{grants} for capability authorization, \texttt{agent\_keys} for signature verification, \texttt{token\_revocations} for invalidated credentials, \texttt{delegations} for credential lineage and revocation, and \texttt{audit\_log} for access evidence. Data payloads remain in backend-native stores and are accessed through adapter-specific query handlers. This control-plane/data-plane separation keeps governance introspectable and auditable without coupling policy state to application data schemas.

\subsection{Multi-Agent Data Exchange and Negotiation}

Data Facts uses a capability-based A2A negotiation protocol (\texttt{negotiation.*} messages/tools).
Agents first discover shareable capabilities, then request authorization, and finally query data through delegated access.

\noindent\textbf{Public:} Capabilities are discoverable and do not require credentials.
Agent~B calls \texttt{negotiate\_discover} and receives public capability entries. If Agent~B calls \texttt{negotiate\_request\_access} for a
public capability, Agent~A returns a public-access acknowledgment (no credential) together with available query names.

\noindent\textbf{Semi-Private:} Capabilities are discoverable but credential-gated.
After discovery, Agent~B sends \texttt{negotiate\_request\_access}. Agent~A either (i) auto-approves and issues a delegated credential (trusted
requester) or (ii) returns a pending request for manual approval.
Agent~B then queries data using \texttt{negotiate\_use\_credential} (A2A proxy), or optionally \texttt{negotiate\_use\_credential\_direct}
(direct gateway mode).

\noindent\textbf{Private:} Capabilities are not discoverable via inquiry.
Agent~B requests a private credential via \texttt{negotiate\_request\_private\_credential} (with a known capability ID). Agent~A issues a short-
lived credential only to trusted agents; untrusted requests are denied.
After credential issuance, Agent~B uses \texttt{negotiate\_use\_credential} to execute queries.

\section{Experiments}
\label{sec:experiments}

We evaluate Data Facts across five experiments targeting the four schema primitives and overall decision quality.
Experiments~\ref{sec:exp-discovery}--\ref{sec:exp-integrity} validate the public-access path; Experiment~\ref{sec:exp-security} validates the private-access pipeline; Experiment~\ref{sec:exp-decisions} measures the downstream effect on agent decision accuracy.

\subsection{Discovery Overhead}
\label{sec:exp-discovery}

\noindent\textbf{Setup.}
We compare two retrieval conditions at $N \in \{50, 200\}$ simulated agent queries.
The \textit{baseline} condition is a single-step direct \texttt{GET} to a known dataset endpoint, representing hard-coded endpoint coupling.
The \textit{treatment} condition executes the full Data Facts path: registry lookup, \texttt{data\_facts\_url} resolution, and dataset fetch (three steps).
The primary metrics are time-to-first-data (TTD) and retrieval success rate.

\noindent\textbf{Results.}
Both conditions achieve 100\% retrieval success at all scales (Table~\ref{tab:discovery}).
The treatment path incurs a fixed overhead of approximately 260--270\,ms, corresponding to two additional HTTP round-trips.
Latency remains stable as $N$ scales from 50 to 200, confirming that overhead is bounded and does not grow with query volume.

\begin{table}[t]
\centering
\caption{Discovery overhead: baseline (direct, 1 step) vs.\
         Data Facts treatment (3 steps).}
\label{tab:discovery}
\small
\setlength{\tabcolsep}{5pt}
\begin{tabular}{llr}
\toprule
\textbf{Condition} & $N$ & \textbf{Mean TTD (s)} \\
\midrule
Baseline  &  50 & 0.120 \\
Treatment &  50 & 0.380 \\
Baseline  & 200 & 0.130 \\
Treatment & 200 & 0.400 \\
\bottomrule
\end{tabular}
\end{table}

\subsection{Freshness Enforcement}
\label{sec:exp-freshness}

\noindent\textbf{Setup.}
A periodically updated dataset is queried under TTL windows of 60\,s, 300\,s, and 600\,s.
Baseline agents apply no staleness check; treatment agents reject data when $t_{\mathrm{now}} - t_{\mathrm{last\_updated}} > \texttt{ttl\_seconds}$.
The outcome metric is decision error rate: the fraction of queries producing an incorrect downstream decision attributable to stale data consumption.

\noindent\textbf{Results.}
We label a request \textit{positive} if the dataset is stale ($\Delta t > \texttt{ttl\_seconds}$). A detection is correct if stale data is rejected. Decision error rate counts cases where stale data is \textit{accepted} (false negatives).
Without Data Facts, the decision error rate is 37.6\% uniformly across all three TTL values: the agent has no mechanism to detect staleness regardless of the configured window.
With Data Facts, the error rate drops to 8.8\% and stale-data detection reaches 76.6\% (confusion matrix: TP\,=\,312, TN\,=\,144, FP\,=\,0, FN\,=\,44).
TTL window size (60--600\,s) has no differential effect; the staleness pattern is independent of window magnitude, and the results are identical across all three TTL conditions.
The binary presence of \texttt{ttl\_seconds} determines whether detection occurs at all.
The residual 8.8\% error represents cases where data is fresh by TTL but has changed within the validity window, an irreducible bound set by the producer's update cadence rather than the schema.

\subsection{Integrity Verification}
\label{sec:exp-integrity}

\noindent\textbf{Setup.}
Payload corruption is injected at rates spanning 1\%--50\%.
Two hundred independent trials are run per corruption level.
Baseline agents ingest payloads without validation; treatment agents recompute the SHA-256 digest and compare it against \texttt{evidence.checksum\_sha256}, rejecting on mismatch.

\noindent\textbf{Results.}
Treatment agents detect 100\% of corrupted payloads at every non-zero corruption level, with zero silent failures.
Baseline agents detect 0\% of corrupted payloads regardless of severity.
The result holds at 1\% corruption: even minimal byte-level modification is reliably caught.
The checksum field provides all-or-nothing detection, not probabilistic detection.

\subsection{Security Pipeline}
\label{sec:exp-security}

\noindent\textbf{Setup.}
A semi-private dataset is stored in PostgreSQL behind a JWT-gated capability gateway (HS256 signing).
Access requires a credential negotiation step that issues a scoped token.
We evaluate four authentication scenarios (valid credentials; missing, empty, and bare-bearer headers), three TTL boundary conditions (immediate use; post-expiry; boundary precision at $t = \mathrm{TTL} \pm \epsilon$), and 46 adversarial attempts across six categories: wrong signing key; six syntactic garbage-token variants; altered \texttt{dataset\_id} claims with valid signatures; impersonation; privilege escalation; and replay (206 total authentication trials; 29 infrastructure-only tests, all $n=20$).

\noindent\textbf{Results.}
Valid-credential access succeeds in 20/20 trials with mean fetch latency 2{,}372\,ms and 100\% SHA-256 checksum verification.
All unauthorized access attempts return HTTP 401 with no unauthorized reads observed in our tests across all authentication scenarios.
TTL enforcement is exact: 45/45 trials classify tokens correctly on both sides of the expiry boundary, with 10/10 correct boundary-precision decisions.
All 46 adversarial attempts are blocked (100\% block rate, 0 data leaks), including altered-claim tokens carrying a valid signature but a modified \texttt{dataset\_id}.
This last result confirms that claim validation is required in addition to signature verification; signature-only JWT validation would pass these tokens.
Gateway JWT verification incurs 48.6\,ms per request, against 259.2\,ms for OAuth2 and 110.5\,ms for SPIFFE mTLS, while enforcing per-dataset capability scoping that API-key schemes cannot provide.

Two behaviors are documented as v1 limitations: token reuse within the validity window is permitted by design in stateless JWT, and tokens are not bound to caller identity.
Both are structural properties of stateless JWT and are addressable via a revocation registry and audience binding in a subsequent revision.

\subsection{Data-Informed Decision Quality}
\label{sec:exp-decisions}

\noindent\textbf{Setup.}
We measure whether the Data Facts access pipeline produces a statistically significant improvement in agent decision accuracy on queries not answerable from LLM training priors.
Seven binary decision scenarios are drawn from a PostgreSQL table with ground-truth distribution Gold\,:\,5, Silver\,:\,6, Bronze\,:\,9.
Each scenario requires exact counts the LLM cannot reliably estimate from priors; the suite spans the full difficulty range from near-certain self-decide success to 0\% (Table~\ref{tab:scenarios}).
Agents operate in two modes: \textit{self-decide} (LLM answers from priors alone) and \textit{data-informed} (LLM queries the database via the Data Facts pipeline before answering).
Three topologies vary infrastructure depth: 1-agent (direct JWT gateway), 2-agent (one A2A credential delegation hop), and 3-agent (two-hop A2A relay chain).
The factorial design yields $7 \times 2 \times 3 \times 20 = 840$ evaluations; all LLM calls use gpt-5-nano.

\begin{table}[t]
\centering
\caption{Decision scenarios ranked by self-decide difficulty (pooled $n=60$ per scenario).
\texttt{near\_parity} is not reliably answerable from priors without database access.}
\label{tab:scenarios}
\small
\setlength{\tabcolsep}{6pt}
\begin{tabular}{lr}
\toprule
\textbf{Scenario} & \textbf{Self-decide (\% correct)} \\
\midrule
\texttt{upgrade\_pool}     & 91.7\% \\
\texttt{tier\_majority}    & 76.7\% \\
\texttt{tier\_gap}         & 50.0\% \\
\texttt{premium\_combined} & 18.3\% \\
\texttt{bronze\_dominance} &  6.7\% \\
\texttt{smallest\_tier}    &  3.3\% \\
\texttt{near\_parity}      &  0.0\% \\
\bottomrule
\end{tabular}
\end{table}

\begin{figure}
    \centering
    \includegraphics[width=1\linewidth]{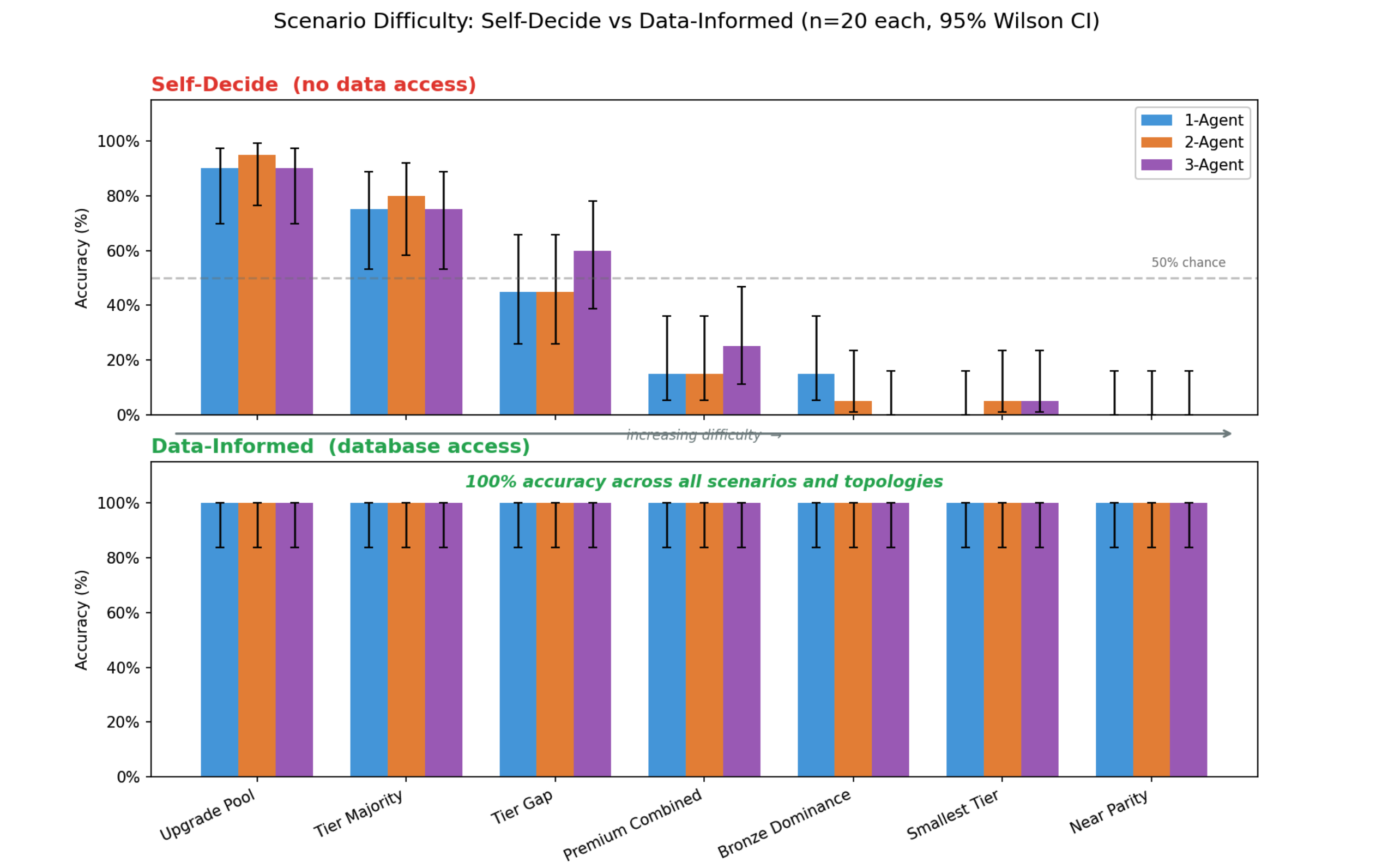}
    \caption{Decision quality on data-driven questions with increasing difficulty.}
    \label{fig:placeholder}
\end{figure}

\noindent\textbf{Results.}
Data-informed agents achieve 100\% accuracy (420/420) across all scenarios and all three topologies, with a pooled 95\% Wilson CI lower bound $\approx99\%$ ($p < 0.001$ against the null hypothesis that data access does not improve accuracy).
Self-decide accuracy is 35.2\% (148/420), 95\% CI [30.9\%, 39.8\%].
Per-topology self-decide accuracy is 34.3\%, 35.0\%, and 36.4\% for 1-, 2-, and 3-agent topologies respectively; the overlapping confidence intervals confirm that topology has no effect on baseline LLM reasoning.
The data-informed access chain succeeds in 420/420 tool-call executions with zero authentication failures, protocol errors, or integrity violations across all three topologies.

The \texttt{near\_parity} scenario achieves 0/60 self-decide accuracy: the question requires $|5-6|=1$, which is inaccessible from training priors regardless of model capability.
It functions as a discriminative control confirming that certain query types are not reliably answerable from priors without database access.

A secondary latency finding (Table~\ref{tab:latency}) shows that data-informed p95 latency is lower than self-decide p95 in all topologies, with the gap widening with infrastructure depth (3-agent: 29.8\,s vs.\ 75.4\,s).
Without ground truth, the LLM incurs deliberation cost on counter-intuitive queries, producing verbose completions that dominate tail latency.
The structured prompt constraining the agent to lead with its answer reduces completion length; the resulting output-token saving outweighs the database round-trip at the p95 tail.

\begin{table*}[t]
\centering
\caption{Latency by topology and mode (seconds).
$\Delta_{\mathrm{mean}}$ is data-informed minus self-decide mean latency (negative indicates data-informed is faster).}
\label{tab:latency}
\small
\setlength{\tabcolsep}{8pt}
\begin{tabular}{lrrrrrr}
\toprule
 & \multicolumn{2}{c}{\textbf{p50 (s)}} & \multicolumn{2}{c}{\textbf{p95 (s)}} & \multicolumn{2}{c}{$\Delta_{\mathrm{mean}}$} \\
\cmidrule(lr){2-3}\cmidrule(lr){4-5}\cmidrule(lr){6-7}
\textbf{Topology} & Self & Data & Self & Data & \multicolumn{2}{c}{} \\
\midrule
1-Agent & 14.8 & 18.2 & 46.8 & 27.7 & \multicolumn{2}{c}{$-1.2$\,s} \\
2-Agent & 25.3 & 17.8 & 52.7 & 23.3 & \multicolumn{2}{c}{$-8.5$\,s} \\
3-Agent & 34.9 & 20.9 & 75.4 & 29.8 & \multicolumn{2}{c}{$-11.6$\,s} \\
\bottomrule
\end{tabular}
\end{table*}

\section{Discussion}
\label{sec:discussion}

The five experiments collectively validate each required field in the Data Facts schema (and by extension the NANDini data exchange layer) and quantify the conditions under which these fields provide measurable benefits for the NANDini data exchange layer.

\noindent\textbf{Schema field necessity.}
The discovery experiment establishes that the external-pointer design imposes bounded, predictable overhead (260\,ms, two HTTP round-trips) independent of query volume and recoverable through TTL-aware caching.
The freshness experiment demonstrates that \texttt{ttl\_seconds} is a necessary field: without it, agents cannot distinguish stale from current data regardless of the window configured, because the check itself cannot be performed.
The residual 8.8\% post-TTL error is irreducible at the schema layer and is set by the producer's update cadence.
The integrity experiment establishes that \texttt{evidence.checksum\_sha256} is not an optional audit aid: silent corruption is a property of the unchecked baseline at every corruption rate, and even 1\% byte-level corruption propagates undetected without the field. We chose SHA-256 over lighter alternatives such as CRC32 or BLAKE3 because its collision resistance properties are well-established in security contexts and its computational cost is negligible relative to the network round-trips already required by the pipeline.
Together, these results confirm the design choice to require freshness and integrity fields rather than mark them optional.

\noindent\textbf{Security design.}
The security results demonstrate that the three-layer pipeline (JWT issuance, gateway claim enforcement, A2A delegation) functions as a strict binary gate.
The salient finding is that claim validation is necessary in addition to signature validation: tokens carrying a valid signature but an altered \texttt{dataset\_id} are correctly rejected, a property that signature-only JWT verification does not provide.
The two v1 limitations (stateless token reuse, absent identity binding) are structural properties of stateless JWT and are independent of the Data Facts schema; both have established mitigations available in the next revision.

\noindent\textbf{Decision quality and topology independence.}
The 0/60 result on \texttt{near\_parity} isolates a class of query that is not reliably answerable from priors without database access, independent of model capability.
The consistency of data-informed accuracy (100\%) and self-decide accuracy (34--36\%) across all three topologies confirms that the A2A credential delegation chain, including the two-hop relay, introduces no information loss and no accuracy penalty relative to direct gateway access.
The latency inversion at p95 suggests a broader principle: on counter-intuitive queries, data access does not merely improve accuracy but also reduces latency by eliminating the LLM deliberation that hard questions induce.

\noindent\textbf{Implications for NANDini.}
These results validate a key premise of the NANDini initiative: that autonomous agents require a structured data layer to make reliable decisions. The observed accuracy gap (100\% vs.\ 35.2\%) demonstrates that agent identity and communication protocols alone are insufficient. Without a mechanism to discover, verify, and access datasets, agents default to reasoning from training priors, leading to unreliable outcomes on data-dependent queries. Data Facts provides the missing link between NANDini's discovery infrastructure (NANDA) and its vision of autonomous, data-driven agent ecosystems operating at planet scale.

\section{Conclusion}
\label{sec:conclusion}

Data Facts, a core concept of the NANDini initiative, addresses a structural gap in current agent infrastructure. While agents can discover one another through NANDA and communicate via A2A and MCP, no machine-readable mechanism currently exists for advertising, evaluating, or verifying the datasets held by participating agents.
Data Facts resolves this gap by introducing a lightweight pointer---added to an existing AgentFacts record---that allows dataset identity, freshness, integrity, and access tier to be resolved on demand. This approach avoids modifications to registry protocols and prevents inflation of registration payloads, thereby operationalizing NANDini's vision of autonomous, data-driven agent ecosystems.
Across 840 decision-making evaluations, data-informed agents achieve 100\% accuracy on counter-intuitive queries where self-deciding agents achieve 35.2\%; TTL enforcement reduces stale-data errors from 38\% to 9\%; SHA-256 checksums detect 100\% of corruption events across all injection rates; and the security pipeline blocks all 46 forgery attempts with zero data leakage.

\noindent\textbf{Limitations.}
All decision-making evaluations use a single model (gpt-5-nano) and a binary classification format; therefore, the results may not generalize to open-ended queries or other model families.
The ground-truth database is static; dynamic environments with high update rates may erode TTL enforcement effectiveness beyond the 8.8\% residual error observed in this study.
Additionally, the v1 security pipeline does not bind tokens to caller identity, allowing intercepted tokens to be reused within their TTL window.

\noindent\textbf{Future work.}
Immediate protocol extensions include token revocation, audience binding, and short-lived credential exchange to address the identity-binding limitation.
The current schema supports only single-dataset documents; introducing a catalog endpoint that lists multiple datasets per agent would reduce round-trips in data-rich deployments.
Within the broader NANDini roadmap, we plan to extend Data Facts along two axes.
First, integration with knowledge distillation pipelines, another NANDini research domain, which would enable agents to advertise not only raw datasets but also distilled knowledge artifacts with associated provenance and quality guarantees.
Second, integration with W3C DCAT federation and PROV lineage fields would enable Data Facts documents to participate in broader data governance workflows without requiring schema changes, advancing NANDini's goal of open, transparent, and planet-scale agent data infrastructure.

\bibliographystyle{plain}
\bibliography{ref}

\end{document}